# Synchrotron X-ray tomography investigation of 3D morphology of intermetallic phases and pores and their effect on the mechanical properties of cast Al-Cu alloys

Yuliang Zhao [a, b, c], Zhi Wang [a, b], Chun Zhang [c, d], Weiwen Zhang [a, b] *

[a] Guangdong Key Laboratory for Advanced Metallic Materials Processing, South China University of Technology, Guangzhou, 510641, China

[b] National Engineering Research Center of Near-net-shape Forming for Metallic Materials, South China University of Technology, Guangzhou, 510641, China

[c] School of Engineering & Computer Science, University of Hull, East Yorkshire, HU6 7RX, UK

[d] School of Mechanics and Civil & Architecture, Northwestern Polytechnical University, Xi'an 710072, China

*Corresponding author: mewzhang@scut.edu.cn (Weiwen Zhang)

**Abstract**

The influence of Fe content on the three-dimensional (3D) morphology of Fe-rich intermetallic phases (Fe phases), $Al_2Cu$, and pores and their effect on the mechanical properties of cast Al-5.0Cu-0.6Mn alloys with 0.5 and 1.0 wt. % Fe are characterized using synchrotron X-ray tomography and a tensile test. The results show that both Fe phases and $Al_2Cu$ exhibit a complex 3D network structure, and the pores are irregular with complex interconnected and near-globular shape. As the Fe content increases from 0.5 % to 1.0 %, the volume fraction and equivalent diameter of Fe phases decrease, whereas $Al_2Cu$ shows the opposite trend; however, both their interconnectivities decrease. Skeletonization analysis of four different Fe phases shows that the Chinese-script-shaped Fe phase is relatively more compact than the plate-like

Fe phases. The equivalent diameter and sphericity of pores vary with Fe content, and their relationships follow exponential functions, $Y = 7.14*X^{-1.29}$ and $Y = 7.06*X^{-1.20}$, respectively. The addition of Fe results in a decrease in the ultimate tensile strength and elongation from 223.7 MPa to 199.8 MPa and from 5.51 % to 3.64 %, respectively, owing to more sharp-edged Fe phases and pores and less $Al_2Cu$, resulting in stress concentration during tensile test.

1. **Introduction**

Aluminum-copper (Al-Cu) alloys are an ideal candidate structure part for iron tools owing to their combination of high strength and toughness [1-2]. Recycled Al-Cu casting alloys are widely used in the transportation sector to reduce vehicle weight, fuel consumption, and greenhouse gas emission [3]. Iron is a common impurity in recycled alloys, and it forms hard and brittle Fe-rich intermetallic phases (called Fe phases hereafter), such as $Al_3(FeMn)$, $Al_6(FeMn)$, $\beta-Al_7Cu_2Fe$, and $\alpha-Al_{15}(FeMn)_3Cu_2$. These phases have detrimental effects on the castability and mechanical properties of the alloys [4]. Mn is a neutral element usually added to Al alloys to transform the Fe phases into less-detrimental Chinese-script-shaped phase [5]. Generally, Fe phases and $Al_2Cu$ are common intermetallic phases in Al-Cu alloys.

Understanding the fundamental relationship between the 3D structures of different phases and their related properties is important for developing new materials. At present, the characterization of different phases is mainly conducted in two dimensions (2D) using optical or electron microscopy [6-7]. However, these techniques are insufficient to fully reveal the complicated three-dimensional (3D) morphologies and spatial structure of intermetallic phases. Owing to the harmful effect of Fe phases, 3D morphologies of Fe phases in Al-Si [8-9], Al-Mg [10], and Al-Cu [11-13] alloys have been characterized using synchrotron X-ray tomography.

Moreover, the nucleation and growth of Fe phases during the solidification of Al alloys have been investigated using 4D (3D + time) synchrotron X-ray tomography [14-17]. Eutectic $Al_2Cu$ is another important phase in Al-Cu alloys; hence, it is necessary to characterize the 3D morphology of this phase. However, the reported literature [18-20] does not fully reveal the 3D morphology of $Al_2Cu$.

Pores are one of the main casting defects in Al alloys [15, 21-26]. The 3D morphology of pores in Al-Cu alloys in a spatial structure is usually an equiaxed and interconnected columnar structure [21]. Pores drastically deteriorate the mechanical properties of alloys, especially the fatigue properties and ductility [24]. Al-Cu alloys have a long freezing temperature range and thus, pores are formed easily in them. Moreover, the morphology and size distribution of pores have a significant influence on the quality of casting alloys. Two types of pores—shrinkage pores and hydrogen gas pores—are usually observed in cast Al alloys [21, 27]. Shrinkage pores are formed owing to inadequate liquid feeding during solidification contraction [27], whereas hydrogen gas pores are mainly formed owing to air entrapment and the release of dissolved hydrogen gas during solidification [28].

Synchrotron radiation X-ray computed tomography (SRXCT) has emerged as a powerful tool for characterizing the 3D morphology of materials, owing to advantages such as its non-destructive process, phase contrast imaging, high spatial resolution, and precise quantitative measurement [29-31]. Owing to the high brilliance of the third-generation synchrotron radiation facility, it is possible to apply SRXCT to investigate the 3D morphology of intermetallic phases and pore defects of metallic materials with micron and sub-micron spatial resolution [29]. Recently, several SRXCT studies [14-17] have been carried out using ex situ and in situ characterization of Al alloys with 3D X-ray tomography. Recently, the 3D morphology of Fe phases in Al-Cu alloys containing low Fe contents (< 0.1 wt.%) has been characterized using SRXCT [11]. It only exhibited 3D interconnected Fe phases and could not provide further information. However, a complete understanding of the 3D morphology of Fe phases, $Al_2Cu$, and pores of recycled Al-Cu alloys and their influence on the mechanical properties of the alloys has yet to be elucidated.

In the present study, we used SRXCT to study the 3D structures and morphologies of the intermetallic phases and pores formed in two Al-5%Cu-0.6%Mn alloys with the addition of 0.5 and 1.0 wt.% Fe (denoted as 0.5Fe and 1.0Fe alloys, respectively, hereafter). Furthermore, the quantitative analysis of Fe phases, $Al_2Cu$, and pores was performed and their effect on the mechanical properties of recycled Al-Cu alloys was evaluated.

## 2. Experiment

### 2.1 Materials

Pure Al ingot (99.9%), Al-20 % Cu master alloy, Al-10 % Mn master alloy, and Al-10 % Fe master alloy were melted to prepare Al-5.0%Cu-0.6%Mn-0.5%Fe (0.5Fe) and Al-5.0%Cu-0.6%Mn-1.0%Fe (1.0Fe) alloys. A clay graphite crucible was used to melt ~5 kg of raw materials inside an electric resistance furnace at 780 °C. The melt was degassed with 0.5 % $C_2Cl_6$ and thereafter, it was stirred and the dross was removed. Subsequently, the melt was cooled to 710 °C to prepare it for casting. The melt was poured into a steel permanent mold preheated to approximately 250 °C. Cylindrical samples (~Ø10 mm × 20 mm) were extracted from the ingot. Samples for tomography scans were machined into cylindrical shape with a size of Ø 2 mm × 5 mm, as described in Ref. [32]. The 2D microstructure was characterized using an FEI Quanta 200 field emission gun scanning electron microscope (SEM).

### 2.2 Synchrotron X-ray tomography

SRXCT experiments were performed at the TOMCAT beamline of the Swiss Light Source, Paul Scherrer Institute (PSI). The experimental setup is shown schematically in Fig. 1a. The detector used was a GigaFRoST camera at PSI. Polychromatic radiation was used to provide the X-rays. Two thousand projections rotate over 180°, providing an exposure time of 7.0 ms for the two samples. The images were recorded by placing the detector at a distance of 180 mm behind the sample. Five hundred projections (16-bit) were reconstructed into a volume of $500^3$ voxels with a voxel size of $(1.62\ \text{micron})^3$. The adjustable magnification was approximately 6.8×. Image J [33] software was used to adjust the contrast between Fe phases and $Al_2Cu$. The 3D segmentation and volume rendering were performed using Avizo Lite v9.0.1 (VSG, France) with a high-performance computer. The 3D bilateral filter was applied to the raw images in order to increase the contrast and reduce noise. This procedure is similar to that described in Ref. [32, 34]. Finally, the primary aluminum, Fe-rich intermetallic phases, $Al_2Cu$, and pores were segmented by using different global threshold values (Pores: 0–10688; $Al_2Cu$: 38815–65535; Fe-rich intermetallic phases: 26439–38814; Al matrix: 10689–26438), as shown in Fig. 1b. The 3D volume renderings of the microstructure are shown in Fig. 1c and the individual segmented phases and pores are shown in Fig. 1d.

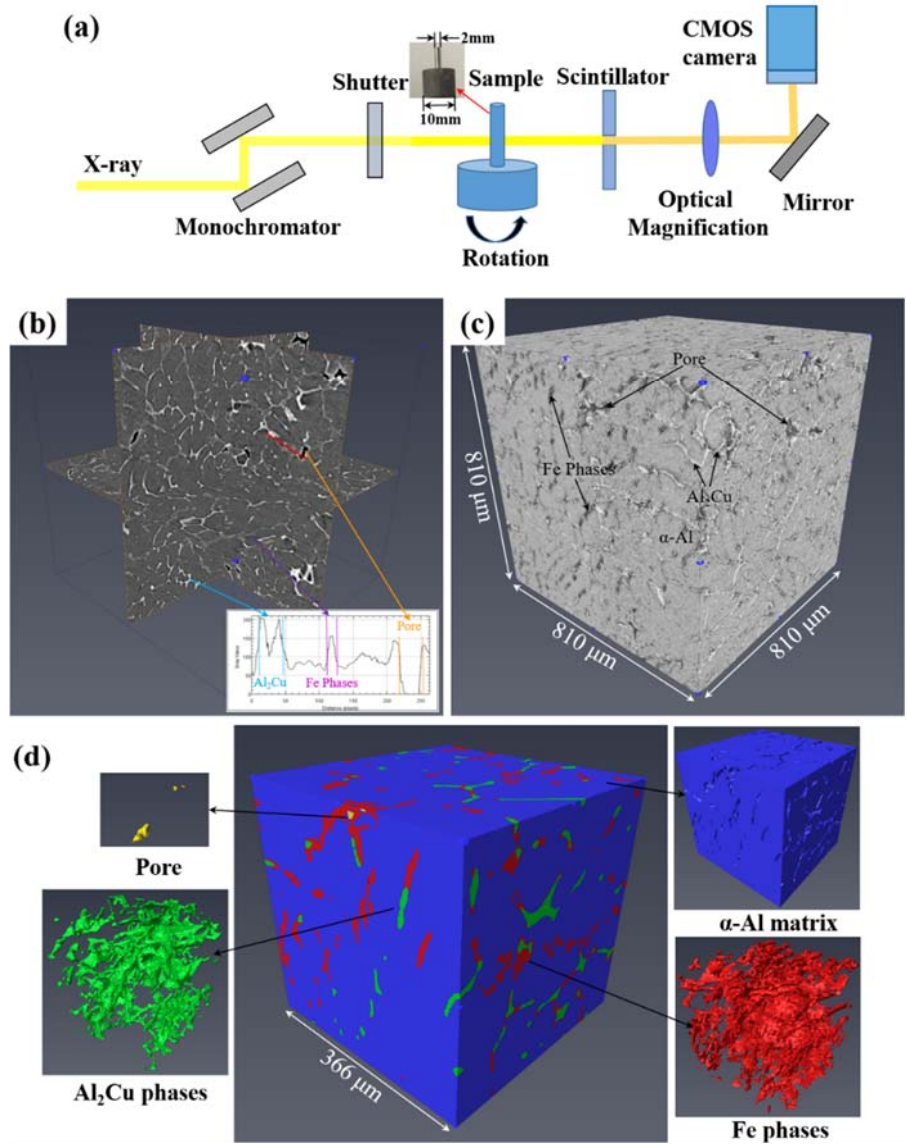

Fig. 1. (a) Schematic diagram of the synchrotron tomography experiment setup at Swiss Light Source, PSI; (b) 3D view of the grey-level distribution in the orthogonal slices of the 0.5Fe alloy (insert: the measured grey values for different phases and pores); (c) 3D volume rendering of the microstructure of the 0.5Fe alloy; (d) the segmented volume rendering of the SRXCT data of the 0.5Fe alloy (blue: Al matrix; red: Fe-rich phases; green: Al$_2$Cu; yellow: pores).

**2.3 Characterization methods**

We used the length, equivalent diameter $D_{eq}$, tortuosity $\tau$, and specific surface area $S_v$ to quantify the size and shape of the Fe phases, and further used the interconnectivity $I$ and mean curvature $H$ to quantify the interconnection of the phases. The equivalent diameter $D_{eq}$ is defined as [21]

$$D_{eq} = \sqrt[3]{\frac{6V}{\pi}}. \tag{1}$$

The specific surface area $S_v$ is defined as the surface-to-volume ratio [8]

$$S_v = \frac{S}{V}. \tag{2}$$

The tortuosity $\tau$ is defined as the geodesic length $L$ divided by the Euclidean length $R$ [27]

$$\tau = \frac{L}{R}. \tag{3}$$

The interconnectivity $I$ is the ratio between the largest 3D individual volume ($V_{larg}$) and the total volume ($V$) [19]

$$I = \frac{V_{larg}}{V}. \tag{4}$$

The mean curvature can be characterized using the two principal radii of curvature, $R_1$ and $R_2$, as follows [20]:

$$H = 0.5 * \left(\frac{1}{R_1} + \frac{1}{R_2}\right). \tag{5}$$

The morphology of each pore was characterized using the following sphericity calculation [35]:

$$\psi = \left(\frac{36\pi V_p^2}{A_p^3}\right)^{\frac{1}{3}}, \tag{6}$$

where $\psi$ is the sphericity of the pore, $V_p$ is the volume, and $A_p$ is the surface area of the pore, where $\psi = 1$ represents a perfect sphere.

## 3. Results and Discussion

Figs. 2(a) and (b) show the SEM images of the 0.5Fe alloy and 1.0Fe alloy, respectively. The microstructure of the as-cast alloys shows that β-Al$_7$Cu$_2$Fe (β-Fe), α-Al$_{15}$(FeMn)$_3$Cu$_2$ (α-Fe), Al$_3$(FeMn), Al$_6$(FeMn), and eutectic Al$_2$Cu phases are randomly distributed in the Al matrix. Figs. 2c and 2d show the tomography scans obtained from SRXCT, illustrating Fe-rich phases, Al$_2$Cu, and pores. A higher fraction of porosity can be observed in the 1.0Fe alloy as compared with that of the 0.5Fe alloy.

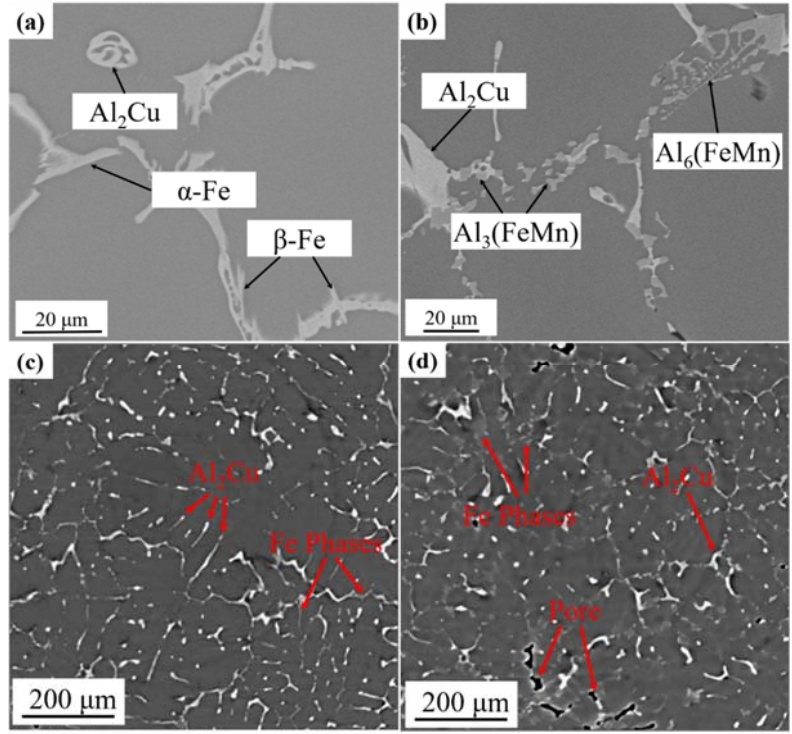

Fig. 2. SEM images of: (a) 0.5Fe alloy, (b) 1.0Fe alloy; the tomography scans of: (c) 0.5Fe alloy, (d) 1.0Fe alloy.

The 3D reconstructed morphologies of Al$_2$Cu and Fe phases in the 0.5Fe and 1.0Fe alloys are shown in Fig. 3. It is evident that the 1.0Fe alloy contains much more disconnected Fe phases (Fig. 3e), whereas the 0.5Fe alloy contains several connected Fe phases (Fig. 3b). The Al$_2$Cu phases extracted from the raw images are shown in Figs. 3c and 3f, illustrating an interconnected network in the 3D structure.

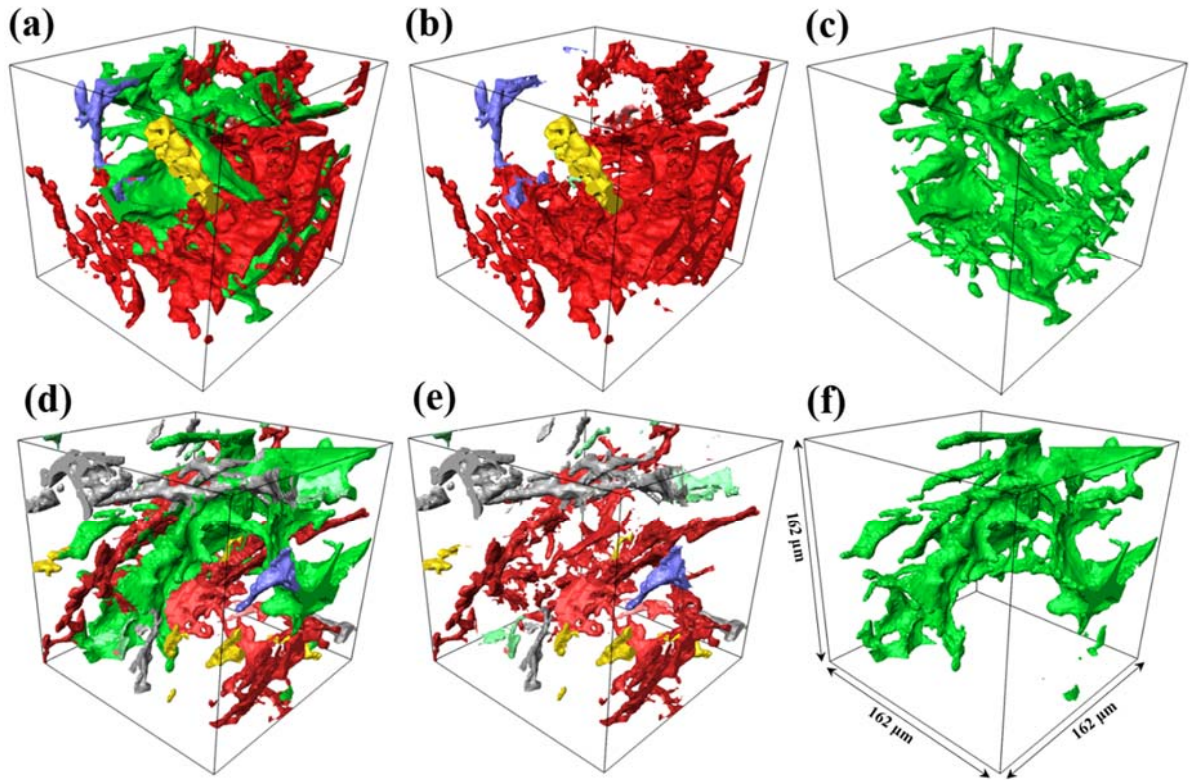

Fig. 3. Subvolume analysis of Fe-rich phases and Al$_2$Cu: (a-c) 0.5Fe alloy; (d-f) 1.0Fe alloy

(Red and other colors: Fe-rich phases; green: Al$_2$Cu).

The 3D reconstructed structures of Fe phases in the 0.5Fe and 1.0Fe alloys in a volume of 810 × 810 × 810 μm$^3$ are shown in Fig. 4. It is observed that Fe phases in the 0.5Fe and 1.0Fe alloys form highly interconnected networks in the 3D structure, which is in accordance with previous reports [11]. The interconnected Fe phases in the 0.5Fe alloy are much fewer than those in the 1.0Fe alloy. Figs. 4c and 4f show that the Fe phases in the 0.5Fe alloy are mainly Chinese-script-shaped α-Fe and a small amount of plate-like β-Fe, whereas they are mainly plate-like Al$_3$(FeMn) and Chinese-script-shaped Al$_6$(FeMn) in the 1.0Fe alloy, which is in accordance with previous reports [34, 36]. The interconnectivity of Fe phases in the 1.0Fe alloy is reduced as less connected plate-like Fe phases are formed in this alloy.

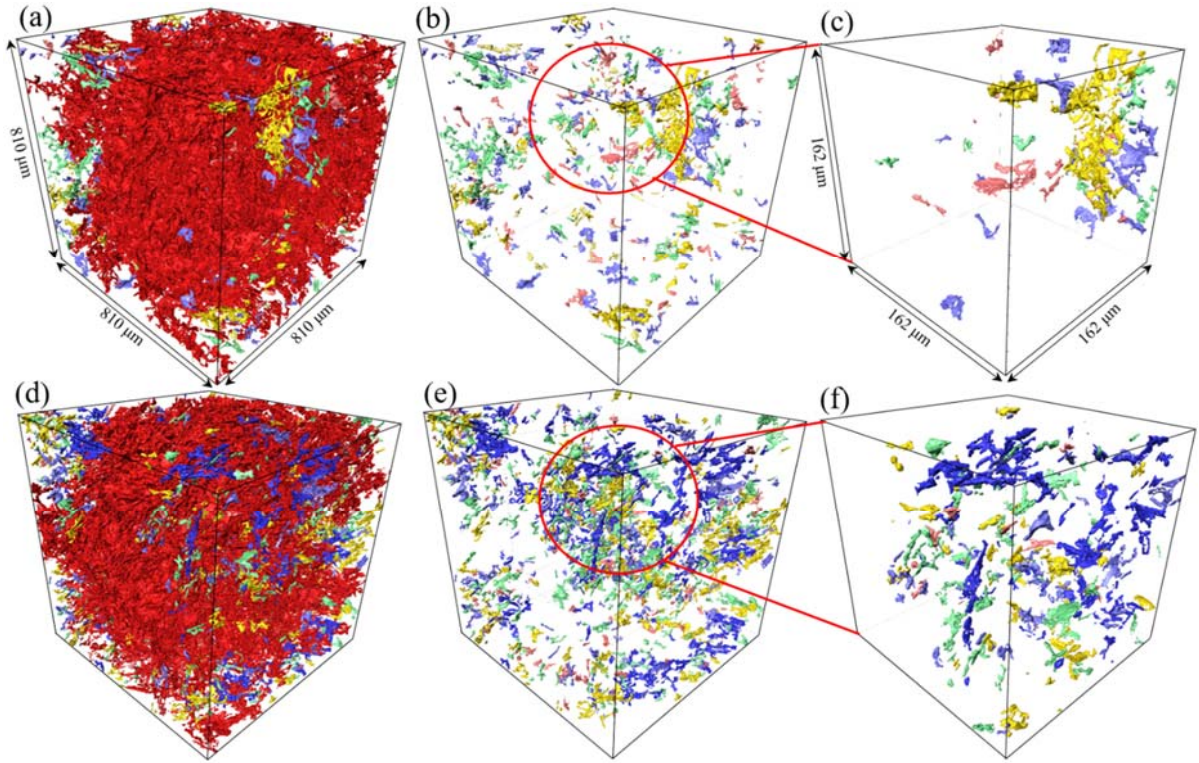

Fig. 4. Volume rendering of Fe-rich phases in alloys in a volume of 810 × 810 × 810 μm³ (a-c) 0.5Fe alloy; (d-f) 1.0Fe alloy (Red: interconnected Fe-rich phases; other colors: separated Fe-rich phases).

The quantitative analysis of Fe phases in the 0.5Fe and 1.0Fe alloys (Fig. 4) is presented in Table 1. The volume fractions of Fe phases are 0.98 % and 1.49 % in the 0.5Fe alloy and 1.0Fe alloy, respectively; their total volumes are 1227376 μm³ and 1862354 μm³, respectively, and the largest particle volumes are 1189327 μm³ and 1651908 μm³, respectively. The equivalent diameters of Fe phases are 132.9 μm and 152.7 μm in the 0.5Fe alloy and 1.0Fe alloy, respectively. The interconnectivity of Fe phases decreases from 96.9 % to 88.7 % when the Fe content in the alloy increases from 0.5 % to 1.0 %. The number of nodes in the 0.5Fe and 1.0Fe alloys is 1677 and 1528, respectively, and the number of segments is 2006 and 1783, respectively. Thus, the quantitative analysis of Fe phases indicated that the equivalent diameter and numbers of nodes and segments of Fe phases increased mainly owing to the increase in the volume fraction of Fe phases. The decrease in the interconnectivity of Fe phases is due to the increase in the volume fraction of the plate-like Fe phase.

**Table 1** Statistics of Fe phases in a volume of 810 × 810 × 810 μm$^3$ and the related skeletonization analysis

| Alloys | Volume fraction (%) | Total Volume (μm$^3$) | Largest particle volume (μm$^3$) | Equivalent diameter (μm) | Interconnectivity (%) | Number of Nodes | Number of Segments |
|---|---|---|---|---|---|---|---|
| 0.5Fe | 0.98 | 1227376 | 1189327 | 132.9 | 96.9 | 1677 | 2006 |
| 1.0Fe | 1.49 | 1862354 | 1651908 | 152.7 | 88.7 | 1528 | 1783 |

In Fig. 5, the volume rendering of the 3D morphology of α-Fe, β-Fe, Al$_3$(MnFe), and Al$_6$(MnFe) phases extracted from the alloys and their skeletonization analysis are displayed separately. The 3D morphologies of different Fe phases are compared with their deep-etched SEM images [34]. Fig. 4a reveals the complex structure of α-Fe and its skeletonization analysis. α-Fe is not typically "Chinese-script-shaped" in the 2D view and shows interconnected complex morphology in the 3D structure. Owing to the low solubility of Fe in molten Al during solidification, Fe atoms accumulated at the liquid–solid interface and promoted a peritectic reaction in the residual melt: L + Al$_6$(FeMn) → α-Al + α-Fe [35]. The 3D morphology of β-Fe is a plate-like shape (Fig. 4b1), which is significantly different from the typical needle-like structure in the 2D view, as previously reported in the literature [14]. β-Fe is formed through the peritectic reaction, L + Al$_3$(FeMn)/Al$_6$(FeMn) → α-Al + β-Fe, at 589–597 °C or the eutectic reaction, L → α-Al + Al$_2$Cu + β-Fe, at 546–547 °C [36, 37]. Fig. 4c1 shows the true 3D morphology of long rod-like Al$_3$(FeMn) no longer appearing "needle-like" as in the 2D morphology. It has a length of ~450 μm and shows a twinning structure with several fine branches. Al$_3$(FeMn) is formed from the liquid through the eutectic reaction, L → α-Al + Al$_3$(FeMn), at 649–653 °C [36]. Fig. 4d1 presents the 3D constructed morphology of Al$_6$(MnFe) with many highly curved and skeleton branches. Although Al$_6$(MnFe) and α-Fe are Chinese-script-shaped in the 2D morphology, in the 3D morphology, α-Fe has an interconnected network and Al$_6$(MnFe) has many branches.

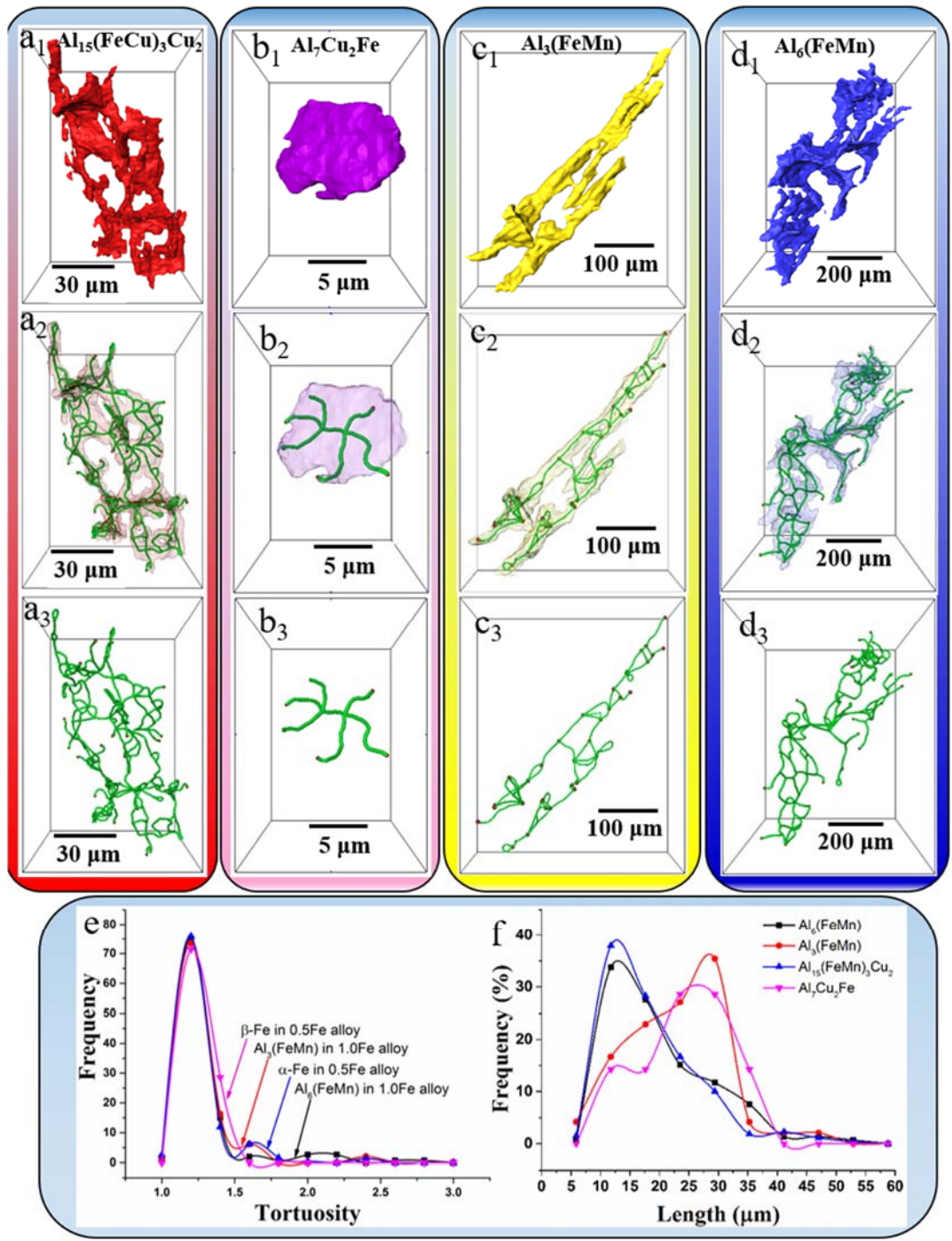

Fig. 5. (a1-d1) Individual 3D morphologies of Fe-rich phases in the alloys [34]; (a2-d3) their skeletonization analyses; (e-f) their quantitative analyses.

The skeletonization function insert in the Avizo® software using the Centerline Tree technique is used to simplify the 3D structures of the phases into 1-voxel-thick skeletons with connected nodes [38]. Thus, the curve length between two nodes and the tortuosity can be calculated, and the 3D morphology of the phases can be quantitatively analyzed. Figs. 5a2-d3 show the skeleton of individual Fe phases, illustrating a complex skeleton structure with many branches. The tortuosity results of Fe phases (Fig. 5e) show that the value distribution is mainly in the range 1.0–1.5. A similar trend in tortuosity suggests that the distribution of Fe phases is random. Skeletonization analysis also provides general information on the distribution of node length for different phases (Fig. 5f). The node length distribution of $Al_6$(FeMn) and α-Fe (5–30 μm) is shorter than that of β-Fe and $Al_3$(FeMn) (7–35 μm), indicating that $Al_6$(FeMn) and α-Fe are relatively compact. This results further confirms that the Chinese-script-shaped $Al_6$(FeMn) and α-Fe are less detrimental than the plate-like β-Fe and $Al_3$(FeMn) during the tensile test.

Fig. 6 presents the typical 3D reconstructions of $Al_2$Cu phases in the 0.5Fe and 1.0Fe alloys. The 3D complex network structure of $Al_2$Cu results from interdendritic lamellar eutectic, which is different from previously reported results [18-20]. In order to illustrate the interconnectivity of $Al_2$Cu phases, the interconnected phases are shown in green, whereas the isolated phases are shown in different colors. It can be observed from Figs. 6c and 6f that the volume fraction of isolated $Al_2$Cu phases in the 1.0Fe alloy is much more than that in the 0.5Fe alloy, indicating that the connectivity of $Al_2$Cu phases in the 1.0Fe alloy is higher than that in the 0.5Fe alloy. Table 2 presents the quantitative statistics data of $Al_2$Cu phases measured from the reconstructed images. The total volume, volume fraction, equivalent diameter, interconnectivity, and numbers of nodes and segments decrease with the increase in Fe content. For example, the volume fraction, equivalent diameter, and interconnectivity decrease from 0.80%, 124.0 μm, and 88.67%, respectively, in the 0.5Fe alloy to 0.63%, 114.8 μm, and 84.65%%, respectively, in the 1.0Fe alloy. These trends are consistent with the explanation provided in the previous reports [7, 35]. The decrease in the volume fraction of Al$_2$Cu with the increase in the Fe content is because more Cu atoms are available in the Fe phases and no more Cu atoms are available for the formation of Al$_2$Cu.

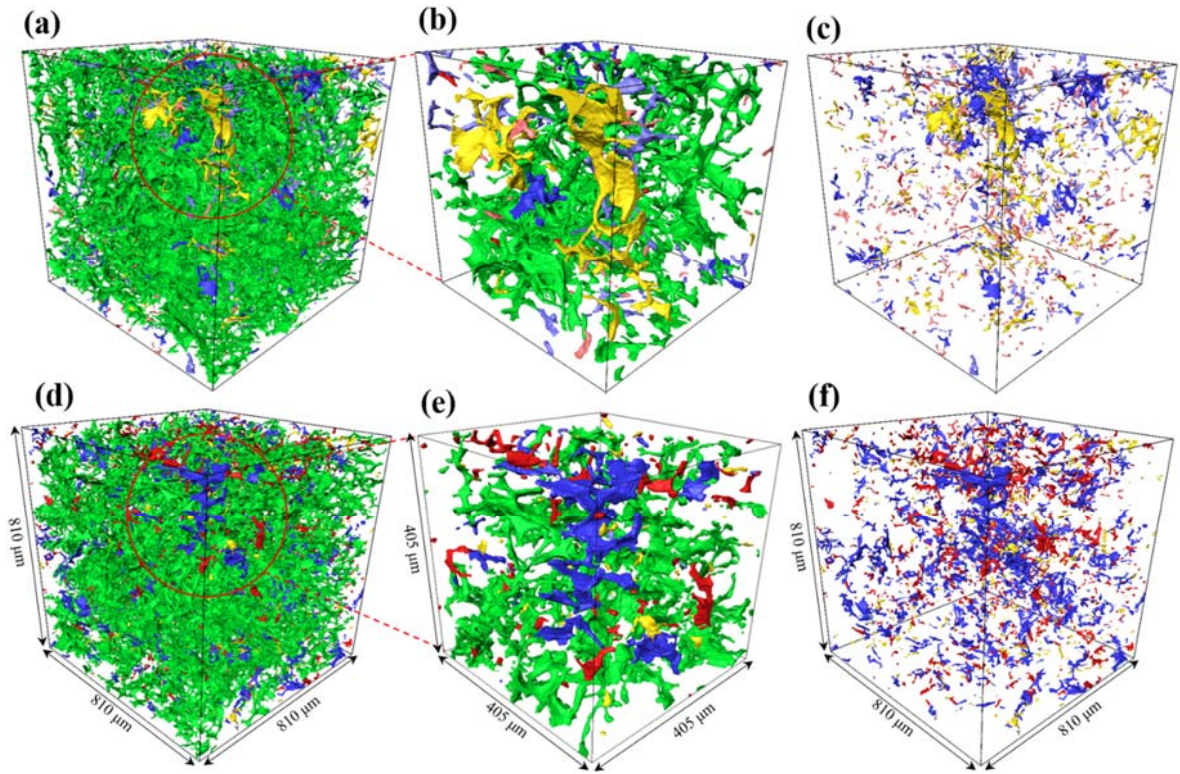

Fig. 6 (a-c) Analysis of Al$_2$Cu in the 0.5Fe alloy in a volume of 810 × 810 × 810 μm$^3$; (d-f) Analysis of Al$_2$Cu in the 1.0Fe alloy in a volume of 810 × 810 × 810 μm$^3$ (Green: interconnected Al$_2$Cu; other colors: separated Al$_2$Cu).

**Table 2** Statistics of Al$_2$Cu phase in a cube with a volume of 810 × 810 × 810 μm$^3$ and the related skeletonization analysis

| Alloys | Volume fraction (%) | Total Volume (μm$^3$) | Largest particle volume (μm$^3$) | Equivalent diameter (μm) | Interconnectivity (%) | Number of Nodes | Number of Segments |
|---|---|---|---|---|---|---|---|
| 0.5Fe | 0.80 | 999395 | 886234 | 124.0 | 88.67 | 1677 | 2006 |
| 1.0Fe | 0.63 | 791831 | 670299 | 114.8 | 84.65 | 1528 | 1783 |

The mean curvature and skeletonization analysis of Al$_2$Cu phase are shown in Fig. 7. Figs. 7a and e show the 3D interconnected complex networks of Al$_2$Cu in the interdendrite of α-Al in the 0.5Fe and 1.0Fe alloys, respectively. The mean curvature of Al$_2$Cu in the 0.5Fe and 1.0Fe alloys is shown in Figs. 7b and f, respectively. Notably, the high positive mean curvature (indicated by red color) is mainly near the α-Al matrix. Al$_2$Cu was a product of a eutectic reaction, and was solidified from the final residual liquid during solidification [24]. The 1.0Fe alloy (Fig. 7f) shows more negative mean curvature regions (blue color) than the 0.5Fe alloy (Fig. 7b). Al$_2$Cu was formed through a eutectic reaction with the previously existing α-Al and Fe phases, and hence, Al$_2$Cu phases in the 1.0Fe alloys contacted with the sharp-edged Fe phases are hollow (negative mean curvature) [34]. The normal distribution of the mean curvature of Al$_2$Cu in the two alloys is shown in Fig. 7i. The asymmetric peak in the normal distribution indicates that the 1.0Fe alloy has a more positive mean curvature owing to the eutectic reaction with sharp-edged Fe phases [34, 36]. The skeletonization analysis of segmented Al$_2$Cu is shown in Figs. 7c, d, g, and h, illustrating typical eutectic networks. The results of the quantitative analysis are presented in Table 3. The total volume and numbers of nodes and segments of Al$_2$Cu decrease from 6013 μm$^3$, 2006, and 1677, respectively, in the 0.5Fe alloy to 5007 μm$^3$, 1783, and 1528, respectively, in the 1.0Fe alloy. The mean radius of Al$_2$Cu is 0.32 μm and 0.31 μm, respectively, and the mean length is 12.52 μm and 12.45 μm, respectively, in the 0.5Fe alloy and 1.0Fe alloy. Thus, Al$_2$Cu in the 1.0Fe alloy is relatively more compact than that in the 0.5Fe alloy.

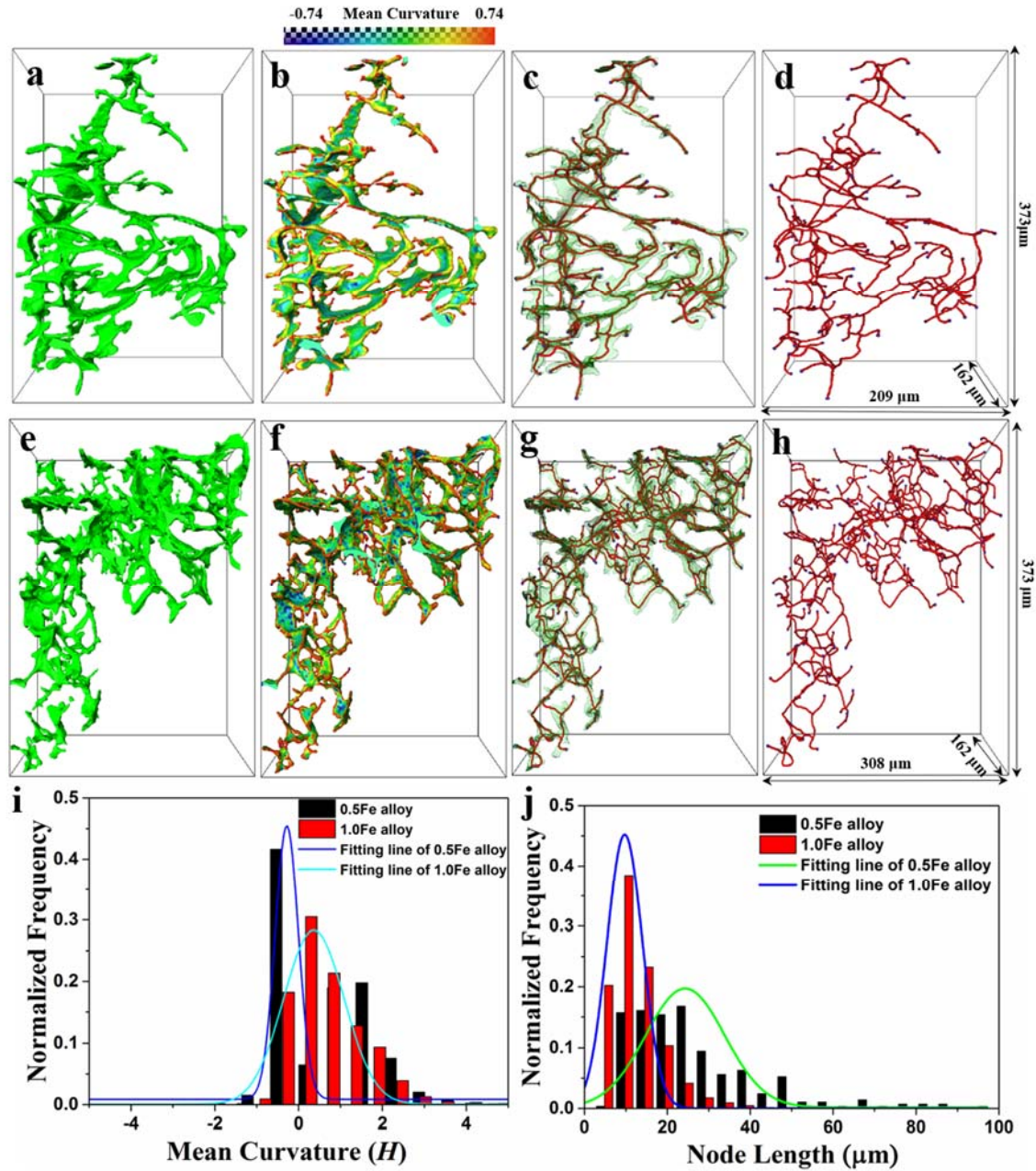

Fig. 7 Typical Al$_2$Cu phase and its quantitative analysis: (a-d) 0.5Fe alloy; (e-h) 1.0Fe alloy; (i-j) Mean curvature and skeletonization analysis of Al$_2$Cu.

Table 3 Summary of quantitative analysis of Al$_2$Cu phase with skeletonization

| Alloys | Total Volume ($\mu m^3$) | Number of Segments | Number of Nodes | Mean Radius | Mean Length ($\mu m$) |
|---|---|---|---|---|---|
| 0.5Fe | 6013 | 2006 | 1677 | 0.32 | 12.52 |
| 1.0Fe | 5007 | 1783 | 1528 | 0.31 | 12.45 |

Pores are a common defect in cast alloys and can limit the fatigue properties and ductility [24] of cast Al alloys. In this study, pores occur as shrinkage pores and gas pores. Quantitative statistics of total volume fraction of pores in 0.5Fe and 1.0Fe alloys are 1.43 % and 2.11 %, respectively. Fig. 8 presents the volume rendering of pores and their sphericity and mean curvature analysis. Numerous equiaxed shrinkage pores and a few near-globular gas pores are observed in both 0.5Fe and 1.0Fe alloys. Shrinkage pores appear to be developed and exhibit a complex interconnected 3D morphology. The low-sphericity feature is evidently constrained by the secondary dendrite arm at the end of solidification. The pore size varies from 30 to 500 μm and the sphericity of the pores is relatively low (0.09–0.63). Gas pores exhibit single near-globular shape with a size of 5–20 μm and sphericity of 0.92–0.95. The mean curvature of the largest shrinkage pore in the 0.5Fe and 1.0Fe alloys is shown in Figs. 8b and d, respectively. It can be observed that pores with high positive mean curvature (red color) are adjacent to the interdendrite region, which is usually contact with the last liquid to solidify. Thus, pores with high negative mean curvature (blue color) are adjacent to the region with primary α-Al [21]. The curvature of pores is influenced by the secondary arm dendrite spacing (SADS) and residual liquid phase when the pore is formed [21]. The two alloys exhibit nearly the same SADS (see Fig. 8) but different volumes of residual liquid phase. As the Fe content is increased, the eutectic temperature increases [36]. The increase in the Fe solute element in the liquid melt and decrease in hydrogen in the Al melt result in a change in the mean curvature of pores [39]. *Much richer and cleaner 3D information and mean curvatures of pores in different view angles are demonstrated in two accompanying videos.*

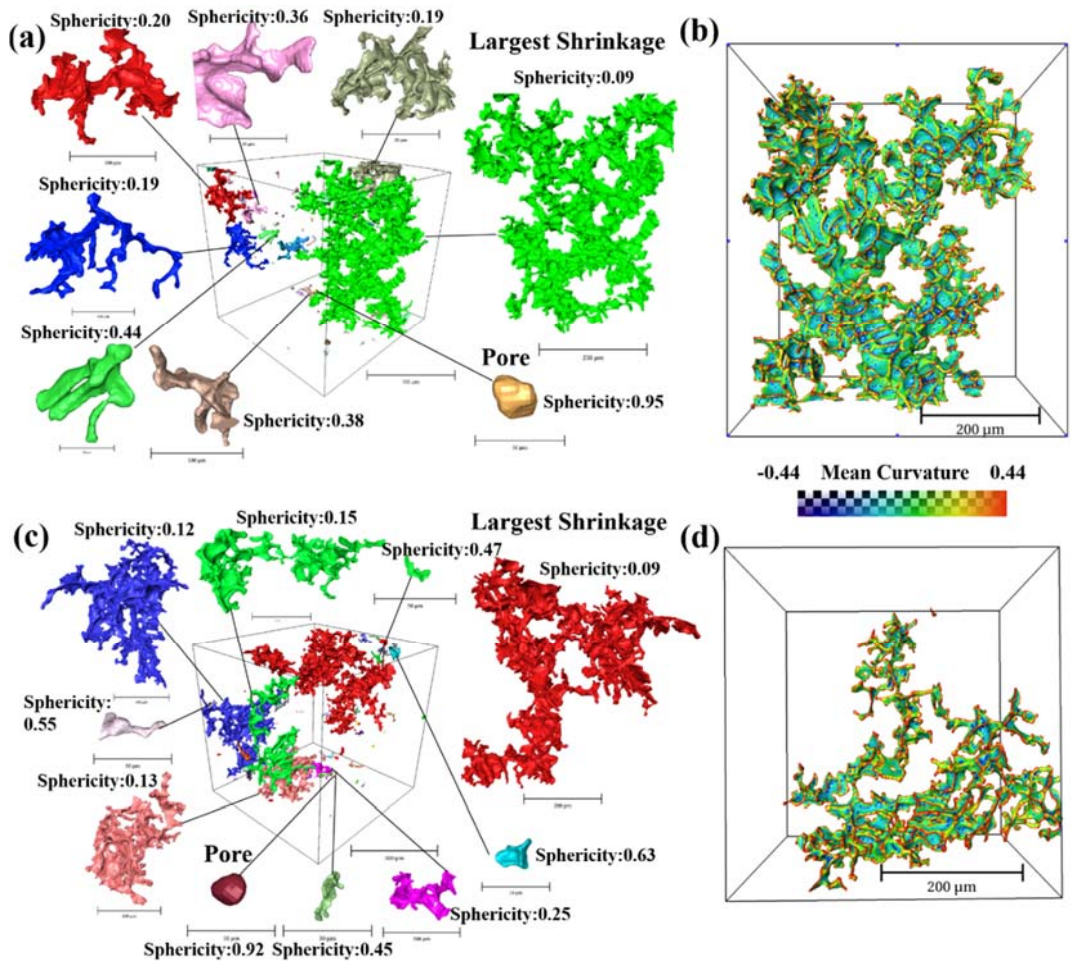

Fig. 8 (a, c) 3D morphology of pores (insert: sphericity) in the 0.5Fe and 1.0Fe alloys; (b, d) the mean curvature of the largest shrinkage pore in the 0.5Fe and 1.0Fe alloys.

Fig. 9 presents the quantitative analysis of pores in the 0.5Fe and 1.0Fe alloys. As shown in Fig. 9a, the equivalent diameter of pores in both alloys is below 80 μm. It is evident that, in the 0.5Fe alloy, the pore size is mostly in the range 5–30 μm, whereas in the 1.0Fe alloy, it is mostly in the range 10–25 μm. Sphericity is a parameter that describes how spherical a pore is [40]. A sphericity of 1 represents a perfect spherical shape. Fig. 9b shows that the sphericity in the 0.5Fe alloy is in the range 0.5–1.0, whereas that in the 1.0Fe alloy is in the range 0.4–1.0. Thus, pores in the 0.5Fe alloy are more spherical. Fig. 9c shows the exponential relationship between the sphericity and equivalent diameter of pores in the 0.5Fe and 1.0Fe alloys; the equations are $Y = 7.14*X^{-1.29}$ and $Y = 7.06*X^{-1.20}$, respectively. Notably, pores with an equivalent diameter greater than 40 μm have a sphericity value between 0.1 and 0.3. This indicates that the greater the equivalent diameter of pores, the smaller the sphericity. The mean curvature of pores follows a normal distribution (Fig. 9d). The distribution peak position (μ) of the pores increases from 0.05 to 0.09 and the standard deviation (σ) decreases from 0.49 to 0.32 as the Fe content increases from 0.5 % to 1.0 %. The difference in the mean curvature of pores is correlated to the gap constrained by the Fe phases, $Al_2Cu$, and Al matrix. $Al_2Cu$ and Fe phases in the 1.0Fe alloy possess high mean curvature; thus, pores in the 1.0Fe alloy possess high mean curvature owing to the pinching effect.

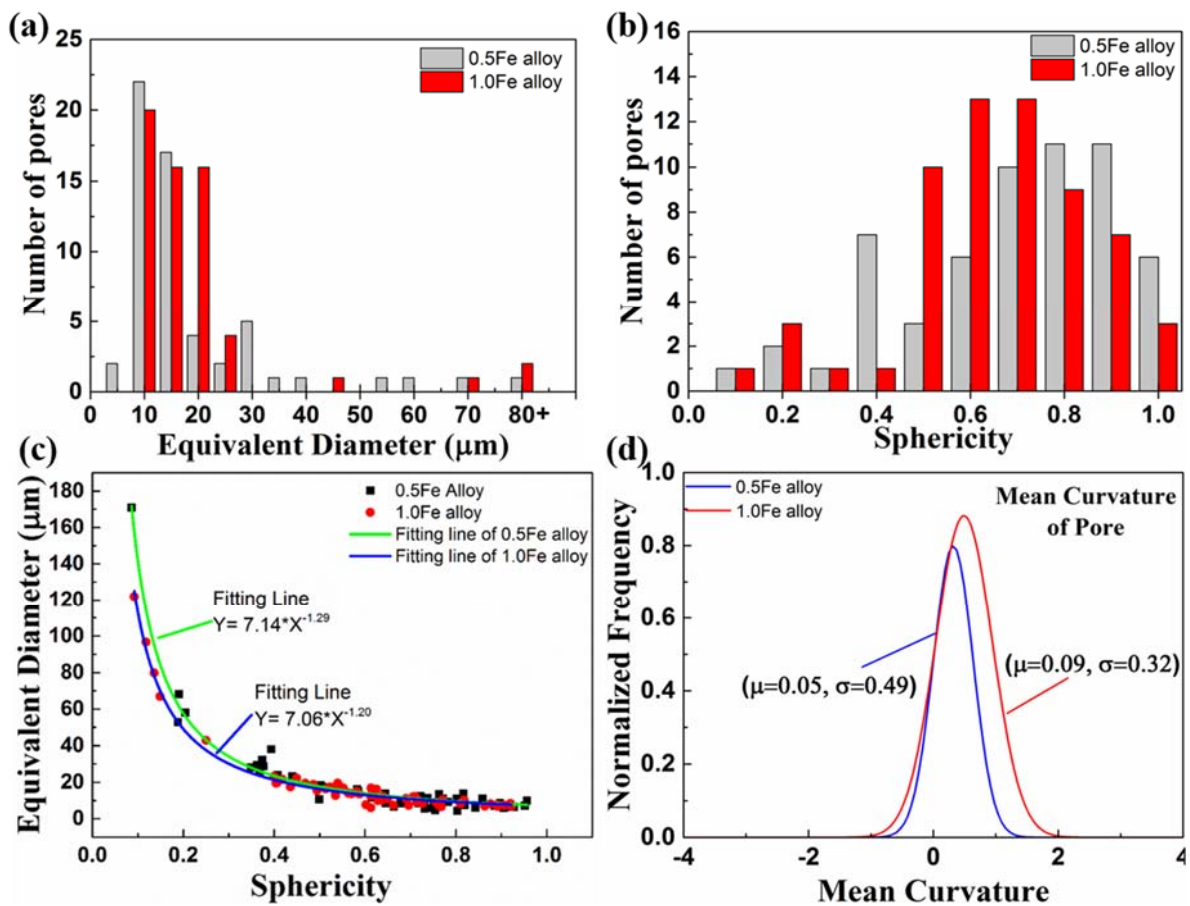

Fig. 9 3D analysis of pores in the 0.5Fe and 1.0Fe alloys: (a-b) size distribution of pores; (c) size distribution of equivalent diameter and sphericity; (d) distribution of mean curvature of pores.

To better understand the influence of intermetallic phases and pores on the mechanical properties of alloys, tensile tests were performed on the 0.5Fe and 1.0Fe alloys. The stress–strain curves of the 0.5Fe and 1.0Fe alloys are shown in Fig. 10. The ultimate tensile strength and elongation of the 0.5Fe alloy are 233.7 MPa and 5.51 %, respectively, whereas the corresponding values for the 1.0Fe alloy are 199.8 MPa and 3.64 %, respectively. The mechanical properties of the alloys decrease as the Fe content increases from 0.5 % to 1.0 %, which is in accordance with previous reports [15, 41].

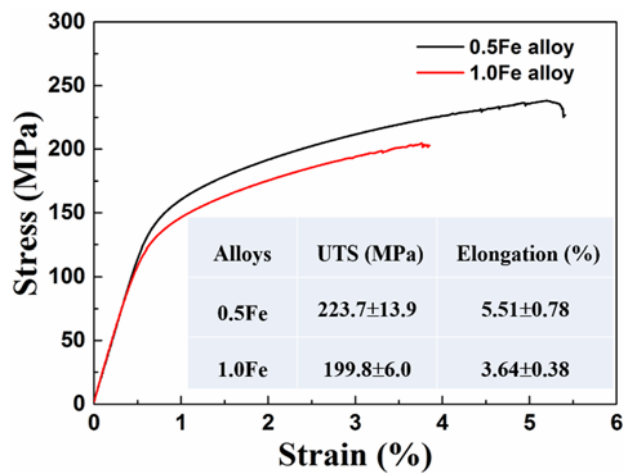

Fig. 10 Stress-strain curves for the 0.5Fe and 1.0Fe alloys

In order to obtain detailed information about the crack initiation and propagation, the post-failure images showing the fracture surfaces of the 0.5Fe and 1.0Fe alloys are displayed in Fig. 11. The 0.5Fe alloy exhibits a rough fracture surface (see Fig. 11a), including pores, dimples, tear ridges, and Fe phases, which indicates that the 0.5Fe alloy experienced a plastic deformation during the tensile test and exhibits semi-ductile surface morphology. The Chinese-script-shaped Fe phases and $Al_2Cu$ can strengthen the alloy by acting as pins to prevent dislocations from sliding under stress. The 1.0Fe alloy exhibits a relatively flat fracture surface (see Figs. 11d-f), which includes pores, tear ridges, and plate-like Fe phases $Al_3(FeMn)$, indicating that the 1.0Fe alloy undergoes brittle fracture. This is further indicated by the flat fracture surface (Fig. 12). The elongated $Al_2Cu$ and Fe phases in the 0.5Fe alloy indicate that it experiences plastic deformation during the tensile test (Fig. 12a), and cracks only exist along the side of plate-like Fe phases (Fig. 12b). Notably, plate-like Fe phases in the 1.0Fe alloy exhibit more damage compared with those in the 0.5Fe alloy because they result in stress localizations, and cracks easily propagate through the Fe phases/Al matrix interface [15, 41].

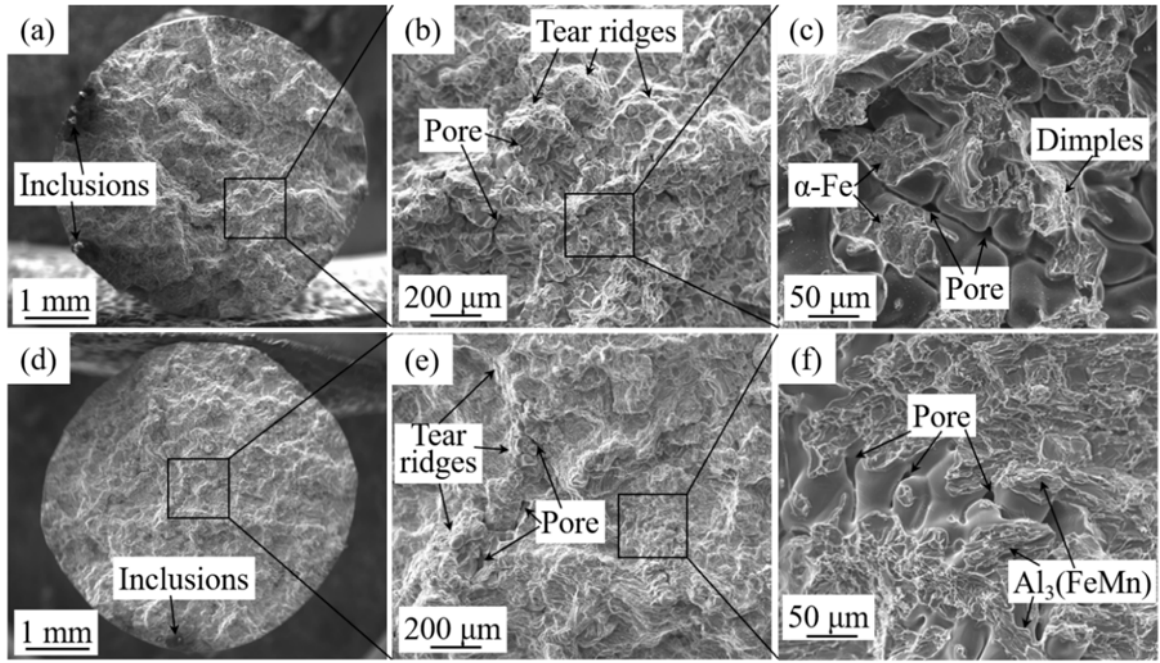

Fig. 11 Fracture surface of different alloys: (a-c) 0.5Fe alloy; (d-f) 1.0Fe alloy.

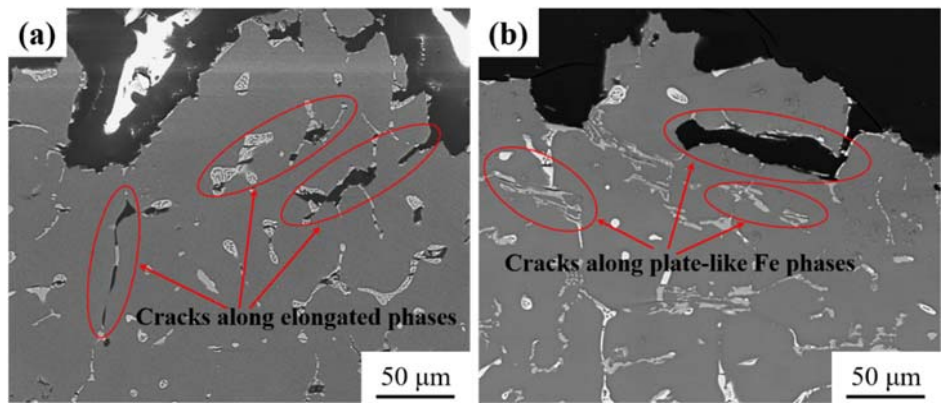

Fig. 12 Flat fracture surface of different alloys: (a) 0.5Fe alloy; (b) 1.0Fe alloy.

The decrease in the ultimate tensile strength and elongation can be explained as follows. First, the volume fraction of sharp-edged Fe phases and pores increased with the increase in Fe content. Pores and intermetallic phases result in stress concentration and act as crack initiation sites [11, 15, 27, 41]. In the case of the 0.5Fe alloy, cracks initiated from hard intermetallic phases were elongated; further, the Chinese-script-shaped Fe phases acted as pins to prevent dislocations from sliding under stress and the alloy experienced plastic deformation. In the case of the 1.0Fe alloy, cracks propagated along the plate-like Fe phases/Al matrix interface, and the alloy experienced brittle fracture. Second, the difference between the hardness of the Al matrix and intermetallic phases resulted in strain incompatibility [42]. According to the nanoindentation measurement [42-44], the hardness values of Fe phases (8.8–14.8 GPa) and the $Al_2Cu$ phase (6.54 GPa) are larger than that of the Al matrix (2.1 GPa). Once cracks have initiated, they propagate along brittle intermetallic phases ($Al_2Cu$ and Fe phases). This can be explained by the different elastic modulus between the Al matrix and intermetallic phases, which provides a preferential path for crack propagation [42].

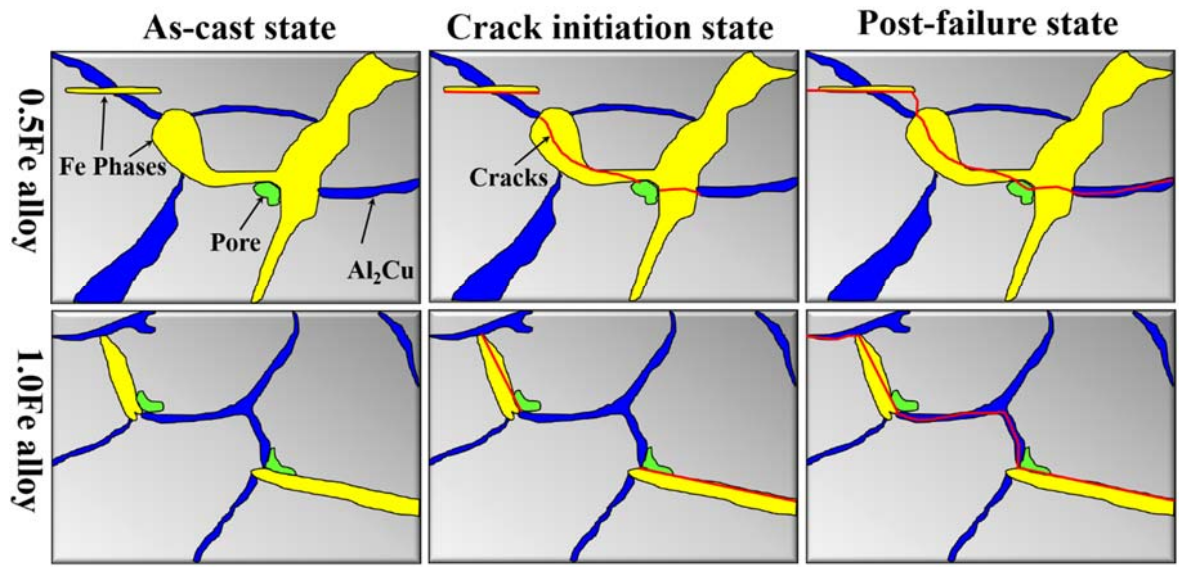

Fig. 13 Schematic illustration of the damage behavior of the alloys. The load direction is vertical.

## 4. Conclusions

In order to study the influence of Fe content on the 3D morphology of intermetallic phases and pores and their effect on the mechanical properties of cast Al-5.0Cu-0.6Mn alloys with 0.5 and 1.0 wt.% Fe, an experiment was performed using SRXCT (at PSI, Swiss Light Source) and a tensile test. The conclusions drawn are as follows:

1. Both Fe phases and Al$_2$Cu exhibited a 3D complex network structure in the alloys. As the Fe content increased from 0.5% to 1.0%, the volume fraction and equivalent diameter of Fe phases decreased, whereas Al$_2$Cu showed the opposite trend. The interconnectivity of both phases increased.

2. Skeletonization analysis of four different Fe phases showed that the Chinese-script-shaped Fe phase was relatively more compact than the plate-like Fe phases. Fe addition increased the positive mean curvature of Al$_2$Cu and the mean length decreased from 12.52 μm to 12.45 μm.

3. Developed and complex interconnected 3D-structured shrinkage pores and near-globular gas pores existed in the alloys. The equivalent diameter of pores in the 0.5Fe and 1.0Fe alloys were mostly in the range 5–30 μm and 10–25 μm, respectively; their sphericity was in the range 0.5–1.0 and 0.4–1.0, respectively. In addition, the relationship between the sphericity and equivalent diameter of pores in the 0.5Fe and 1.0Fe alloys followed the exponential functions $Y = 7.14 \ast X^{-1.29}$ and $Y = 7.06 \ast X^{-1.20}$, respectively.

4. As the Fe content increased from 0.5% to 1.0%, the ultimate tensile strength decreased from 223.7 MPa to 199.8 MPa and elongation decreased from 5.51 % to 3.64 % because more sharp-edged Fe phases resulted in stress concentration during the tensile test.

**Acknowledgement**

Authors gratefully acknowledge the support from Natural Science Foundation of China (51374110 and 51701075), and Team project of Natural Science Foundation of Guangdong Province (2015A030312003). The authors wish to thank all staff member of TOMCAT beamtime of Swiss Light Source, Paul Scherrer Institute. They also grateful to Professor Jiawei Mi's group at Hull University for providing excellent working and studying environment and High-Performance Computing facility. Financial support from the Chinese Scholarship

Council (for Yuliang Zhao's PhD study at Hull University in Nov. 2016 - Nov. 2017) is also acknowledged.**References**

[1] W.C. Harrigan, Handbook of metallic composites, Marcel Dekker, New York, 1994.

[2] J. Brown, Foseco non-ferrous foundryman's handbook, Butterworth-Heinemann, 1999.

[3] J.A.S. Green, Aluminum recycling and processing for energy conservation and sustainability, ASM International, Materials Park, OH, 2007.

[4] L.F. Zhang J.W. Gao, L.N.W. Damoah, D.G. Robertson, Removal of iron from aluminum: a review, Mine. Process. Extr. M. 33 (2012) 99-157.

[5] W.W. Zhang, B. Lin, J.L. Fan, D. T. Zhang, Y.Y. Li, Microstructures and mechanical properties of heat-treated Al-5.0Cu-0.5Fe squeeze cast alloys with different Mn/Fe ratio, Mater. Sci. Eng. A 588 (2013) 366-375.

[6] Y.L. Zhao, W.W. Zhang, C. Yang, D.T. Zhang, Z. Wang, Effect of Si on Fe-rich intermetallic formation and mechanical properties of heat-treated Al-Cu-Mn-Fe alloys, J. Mater., Res. 2018. DOI: 10.1557/jmr.2017.441.

[7] B. Lin, W.W. Zhang, Y.L. Zhao, Y.Y. Li, Solid-state transformation of Fe-rich intermetallic phases in Al-5.0Cu-0.6Mn squeeze cast alloy with variable Fe contents during solution heat treatment, Mater. Character. 104 (2015) 124-131.

[8] Z. Asghar, G. Requena, H.P. Degischer, P. Cloetens, Three-dimensional study of Ni aluminides in an AlSi12 alloy by means of light optical and synchrotron microtomography, Acta Mater. 57 (2009) 4125-4132.

[9] A. Tireira, G. Requena, S.S. Jao, A. Borbely, H. Klocker, Rupture of intermetallic networks and strain localization in cast AlSi12Ni alloy: 2D and 3D characterization, Acta Mater. 112 (2016) 162-170.

[10] D. Tolnai, G. Requena, P. Cloetens, J. Lendvai, H.P. Degischer, Effect of solution heat treatment on the internal architecture and compressive strength of an AlMg4.7Si8 alloy, Mater. Sci. Eng. A 585 (2013) 480-487.

[11] R. Fernández Gutiérrez, F. Sket, E. Maire, F. Wilde, E. Boller, G. Requena, Effect of solution heat treatment on microstructure and damage accumulation in cast Al-Cu alloys, J. Alloys Compd. 697 (2017) 341-352.24